\documentclass[%
reprint,
showpacs,preprintnumbers,
amsmath,amssymb,
longbibliography
]{revtex4-2}

\usepackage{graphicx}
\usepackage{dcolumn}
\usepackage{bm}
\usepackage{colortbl}
\usepackage[unicode=true,colorlinks=true,citecolor=blue,urlcolor=blue]{hyperref}
\usepackage{braket}
\usepackage[normalem]{ulem}

\renewcommand{\Im}{\mathop{\rm Im}}

\newcommand{\e}{\mathrm{e}}

\let\ifr\i
\renewcommand{\i}{{\rm i}}
\renewcommand{\d}{\mathrm d}
\renewcommand{\emph}{\textit}
\newcommand{\aver}[1]{\left \langle #1 \right \rangle}

\begin{document}

\title{Intervalley mixing of interface excitons at lateral heterojunctions}

\author{M.~V.~Durnev}
\affiliation{Ioffe Institute, 194021 St. Petersburg, Russia}
\author{D.~S.~Smirnov}
\email[Electronic address: ]{smirnov@mail.ioffe.ru}
\affiliation{Ioffe Institute, 194021 St. Petersburg, Russia}

\begin{abstract}
 We demonstrate that the low symmetry of armchair lateral heterojunction between transition metal dichalcogenide monolayers allows for the mixing of $\bm K_+$ and $\bm K_-$ valleys. From the tight binding model we estimate the strength of the valley coupling to be of the order of $0.2$~eV$\cdot$\AA~for typical heteropairs. We show that the valley mixing gives rise to the in-plane $g$-factor of localized electrons leading to the spin precession in the in-plane magnetic field. We further study the effects of the valley mixing on the fine structure and dynamics of excitons at type-II lateral heterojunctions. We find that the interplay of the valley mixing and long-range exchange interaction leads to the linear polarization of exciton photoluminescence along the armchair heterojunction with the degree of polarization up to $1/3$ under unpolarized excitation. Application of the in-plane magnetic field of the order of 10--100~mT in any direction leads to the depolarization of the photoluminescence.
\end{abstract}

\maketitle{}

\section{Introduction}
\label{sec:intro}

Vertical stacks of transition metal dichalcogenides (TMDCs) and other two-dimensional (2D) crystals have attracted enormous attention in recent years~\cite{Geim:2013,Novoselov:2016,Liu:2016,Jin:2018,Castellanos-Gomez:2022}.
Their optical properties are determined by intra- as well as interlayer excitons and trions~\cite{Durnev:2018,Wang:2018,Jiang:2021,Holler:2024,Brotons-Gisbert:2024}.
Due to the type-II band alignment between different TMDCs~\cite{Chiu:2015,Guo:2016,Davies:2021}, electron and hole in interlayer exciton are separated in vertical direction and are bound by the Coulomb interaction in lateral directions.
The incommensurability of the lattice constants of different monolayers as well as tunable interlayer twist lead to the alternation of the local atomic registers between monolayers~\cite{Kang:2013,Yu:2017,Tran:2020}. The resulting moir\'e potential (including possible lattice reconstruction~\cite{Rosenberger:2020,Weston:2020}) gives rise to a whole branch of exciting phenomena including moir\'e excitons~\cite{urbaszek2019materials}, circular dichroism~\cite{PhysRevB.104.L241401,Michl2022,Wang2022}, moir\'e phonons~\cite{doi:10.1021/acsnano.8b05006,Parzefall_2021}, and exotic many-body electronic phases~\cite{Wilson:2021,Regan:2022,Mak:2022}. Even reacher physics is expected in few layer vertical heterostructures~\cite{Tong_2020,brotons2020spin,Zhang:2023a}.

\begin{figure}[htpb]
\begin{center}
  \includegraphics[width=0.9\linewidth]{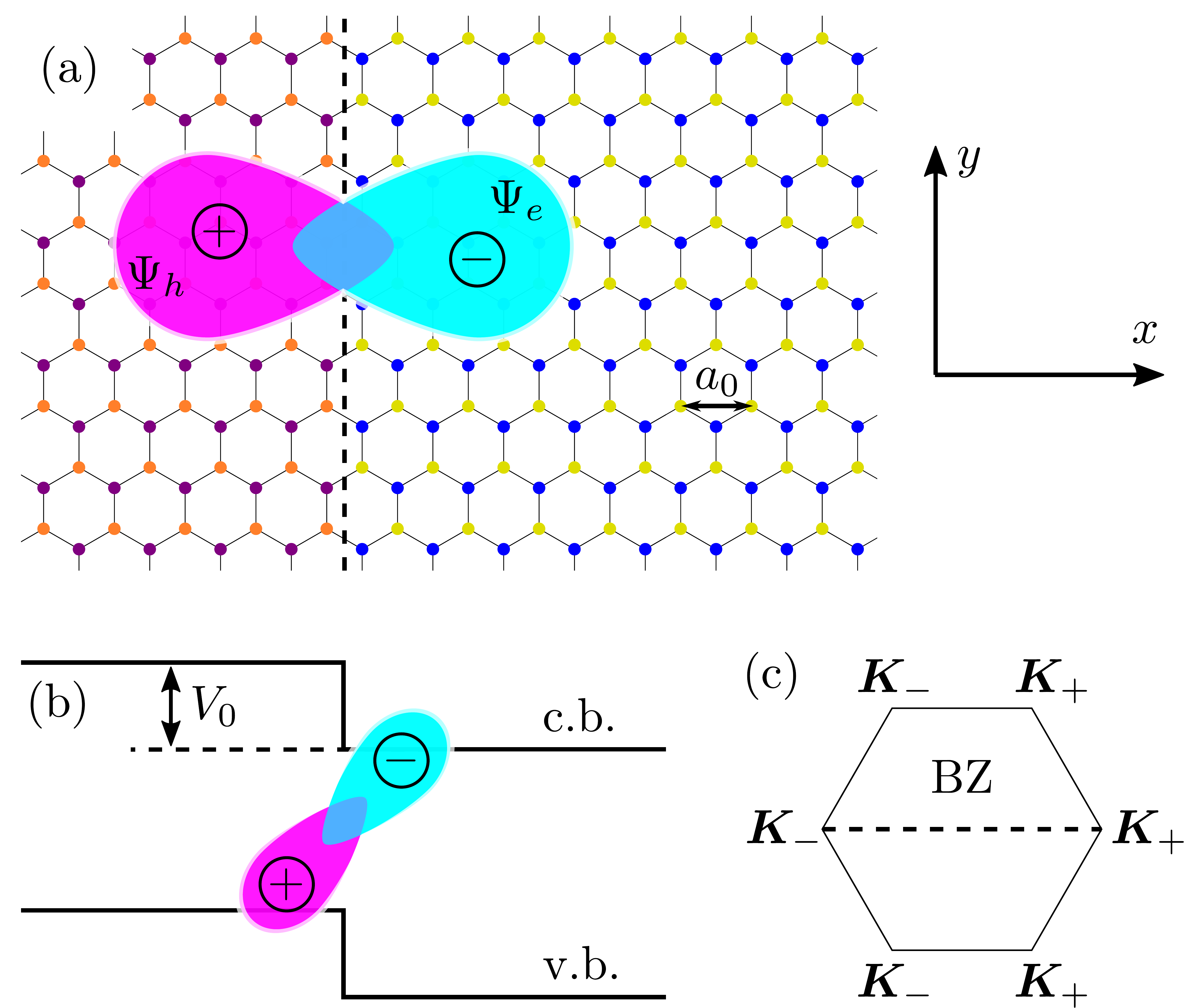}
 \end{center}
  \caption{\label{fig:system}
    (a) Sketch of an interface exciton localized at the armchair lateral heterojunction. (b) Type-II band alignment of conduction and valence bands resulting in an electron-hole separation in the direction normal to the heterojunction. (c) Orientation of the Brillouin zone of transition metal dichalcogenide monolayer. 
  }
\end{figure}

A promising emerging class of 2D materials is associated with the lateral heterostructures,
where different TMDC monolayers are stitched together forming one-dimensional interfaces~\cite{Li:2016,Avalos-Ovando:2019a,Swain:2021,Castellanos-Gomez:2022}. Such structures are typically grown in a shape of triangular quantum dots by chemical vapor deposition. In the growth process, a core of a quantum dot made of one constituent is surrounded by a shell of another constituent~\cite{Duan:2014,Huang:2014,Heo:2015,Ullah:2017}. Single- and multi-junction lateral heterostructures with atomically sharp interfaces have been synthesized~\cite{Li:2015,Sahoo:2018,Ichinose:2022}.
Similarly to vertical heterostructures, electron and hole at a lateral heterojunction are separated due to the type-II band alignment, but the Coulomb interaction can still bind them together, as shown in Fig.~\ref{fig:system}. In this way, spatially indirect interface excitons form~\cite{Lau:2018}. The signatures of such interface excitons in the emission spectra of lateral heterojunctions have been recently reported in Refs.~\cite{Rosati:2023,Yuan:2023}. 

The spatial separation of electron and hole increases the exciton radiative lifetime on the one hand and creates a giant static dipole moment perpendicular to the interface on the other hand~\cite{Yuan:2023}. As a result, lateral heterojunctions represent a promising platform for the realization of strongly interacting and correlated one dimensional states of long-living interface excitons~\cite{Thureja:2022,Avalos-Ovando:2019a}.

Precise control of excitons requires detailed knowledge of their fine structure. For interface excitons, it is determined by the translational symmetry breaking at the interface. In particular, this symmetry breaking leads to the valley mixing at the armchair heterojunctions, since the two valleys are projected to the same point of the one-dimensional Brillouin zone in this case~\cite{Liu:2014,Rostami:2016,Avalos-Ovando:2019}, Fig.~\ref{fig:system}(c). In this work, we elaborate on the important physical effects caused by this mixing, such as the intervalley Zeeman splitting of electron states and polarized photoluminescence of lateral heterojunctions in external in-plane magnetic field. We also present an analytical theory for the valley-mixing constant based on the tight binding (TB) model and calculate the fine structure of interface excitons localized at atomically sharp lateral heterojunctions.

\section{Interface-induced valley mixing}
\label{sec:mixing}

Consider an electron localized at the lateral heterointerface between two different $MX_2$ compounds, where $M$ and $X$ are the metal and chalcogen atoms, respectively, Fig.~\ref{fig:system}(a). Just as in the case of the edges in graphene, a particular armchair configuration of the heterointerface in $MX_2$ heterostructure results in the coupling of the $K_+$ and $K_-$ valleys~\cite{Brey:2006,Akhmerov:2008,Liu:2014,Peterfalvi:2015,Rostami:2016,Avalos-Ovando:2019}. Hence, the
electron wave function for the armchair interface can be presented as
\begin{equation}
\label{envelope}
\Psi_e(\bm r) =  \varphi_1(x) \e^{\i K x} u_{K_+} (\bm r) + \varphi_2(x) \e^{-\i K x} u_{K_-} (\bm r)\:,
\end{equation}
where $u_{K_\pm}$ are the periodic Bloch amplitudes of the conduction-band minima at the $\bm K_\pm = (\pm K, 0)$ points, $\varphi_1$ and $\varphi_2$ are the smooth (in general, two-component) functions, $K = 4 \pi/3a_0$, $a_0$ is the lattice constant, and $\bm r = (x,y)$. 

Intervalley mixing at the armchair interface parallel to the $y$ axis and located at the coordinate $x_{\rm int}$ along the $x$ axis can be described by the effective Hamiltonian
\begin{equation}
\label{Hval}
\mathcal H_{\rm val} = \lambda \e^{-2\i K x_{\rm int}}\tau_+ \delta(x - x_{\rm int})+{\rm H.c.}\:,
\end{equation}
where $\lambda$ is the (complex) valley-mixing constant,  $\tau_+=(\tau_x+\i\tau_y)/2$ is the valley raising operator expressed through the components of the valley Pauli matrices $\bm\tau = (\tau_x, \tau_y, \tau_z)$ with $\tau_z=\pm1$ corresponding to $K_\pm$ valleys, respectively, $\delta(x)$ is the Dirac delta-function and ${\rm H.c.}$ stands for the Hermitian conjugate.
The envelope function approximation with the valley-mixing term~\eqref{Hval} has been applied to calculate the valley splitting in Si/SiGe quantum wells~\cite{Nestoklon:2006}. For a TMDC-based lateral heterojunction, Eq.~\eqref{Hval} describes the main contribution to the intervalley mixing, which is spin conserving and hence most pronounced for the conduction-band electrons due to their smaller spin splitting~\cite{Liu:2014,Rostami:2016}.

The valley-mixing constant $\lambda$ can be derived from a simple one-band TB model with single $d_{z^2}$ orbitals at metal atoms, which give the dominant contribution to the Bloch functions $u_{K_\pm}$ in the conduction band~\cite{Liu:2013, Shi:2013}. The corresponding sublattice is trigonal and has one atom per unit cell. In the following we set the interface coordinate $x_{\rm int} = 0$, so that the metal atoms are located at coordinates $x_n=a_0/2(n+1/2)$ with integer $n$. 
Since $\lambda$ does not depend on the electron wave vector along the heterojunction, we set this vector to zero. Hence, the amplitudes of the TB wave function $c_n$ satisfy the following Schr\"odinger equation
\begin{equation}
\label{tb}
(E - V_n) c_n = t \left(c_{n-2} + 2 c_{n-1}+ 2 c_{n+1} + c_{n+2} \right)\:,
\end{equation}
where $E$ is the energy, $t$ is the hopping matrix element assumed to be identical for both materials (resulting in equal effective electron masses), and $V_n$ is the electron potential energy at the $n$th atom. We set $V_n = V_+$ for the metal atoms to the right side of the interface ($n = 0,1,\dots$), and $V_n = V_-$ for the metal atoms to the left side of the interface ($n = -1,-2\dots$).
The effective electron mass at the $K$-points in this model is $m_c=2\hbar^2/(3ta_0^2)$.

In the envelope function approximation, the TB wave function can be written as
\begin{equation}
\label{cn}
c_n = u(x_n) \e^{\i K x_n} + v(x_n) \e^{-\i K x_n}\:,
\end{equation}
where $u(x)$ and $v(x)$ are the smooth envelop functions in the $K_\pm$ valleys, respectively. The boundary conditions for $u(x)$ and $v(x)$ at $x = 0$ are then derived from Eq.~\eqref{tb}, see also Refs.~\cite{Ando:1982,Ivchenko:1996} for a similar approach.
For realistic structures $t\gg|V_+-V_-|$, so that $u$ and $v$ can be decomposed in the Taylor series to the left and to the right from the heterojunction as
\begin{equation}
\label{decompose}
u(x) \approx \sum_\pm \left[ u_\pm(0) + u_\pm'(0) x + u_\pm''(0) \frac{x^2}{2} \right] \Theta(\pm x)\:.
\end{equation}
Here, $u_\pm(0)$, $u'_\pm(0)$ and $u''_\pm(0)$ are the functions, their first and second derivatives over $x$ taken at $x$ approaching the interface from the right- ($+$ subscript) and left-hand ($-$ subscript) sides, respectively. 
In the last term, the second derivative can be substituted from the Sch\"odinger equation neglecting the valley mixing:
\begin{eqnarray}
\label{d2u}
u_\pm''(0) = \frac{2m_c(V_\pm - E - 3t)}{\hbar^2} u_\pm(0)\:.
\end{eqnarray}

The set of Eqs.~\eqref{tb} for $n=-2,-1,0,1$ yields the four boundary conditions for the envelope functions: the 
continuity of the functions itself, $u_+(0) = u_-(0),~v_+(0) = v_-(0)$, and the discontinuity of their derivatives given by
\begin{eqnarray}
\label{bc}
u_+'(0) - u_-'(0) &=& -\i \alpha \left[ u(0) + v(0) \right] \:, \\
v_+'(0) - v_-'(0) &=& \i \alpha \left[u(0) +  v(0) \right] \nonumber\:
\end{eqnarray}
with
\begin{equation}
\label{alpha}
\alpha = \frac{(V_+ - V_-) a_0 m_c}{\sqrt{3} \hbar^2}\:.
\end{equation}
The boundary conditions Eq.~\eqref{bc} lead to the valley mixing, since the discontinuity of $u'(0)$ is proportional to $v(0)$ and vice versa~\footnote{The intravalley contributions to the discontinuity of the derivatives, i.e. $u_+'(0) - u_-'(0) \propto u(0)$ and $v_+'(0) - v_-'(0) \propto v(0)$ are related to the cubic in $k_x$ terms in the energy dispersion of $K_\pm$ valleys, $E_{K_\pm} \approx - 3t + \hbar^2 (k_x^2+k_y^2)/2m_c \mp \hbar^2 a_0 (k_x^3 - 3k_x k_y^2)/4\sqrt{3}m_c$}.
These boundary conditions are equivalent to adding the $\delta$-functional term $\mathcal H_{\rm val}$, Eq.~\eqref{Hval}, in the Schr\"odinger equation for the envelopes $\varphi_1$ and $\varphi_2$~\cite{Ivchenko:1996,Nestoklon:2006} with
\begin{equation}
\label{lambda}
\lambda = -\i \frac{(V_+ - V_-) a_0}{2\sqrt{3}}\:.
\end{equation}
Note that this result can be also obtained by calculating the average of the electron potential with the TB wave function~\eqref{cn}~\cite{Saraiva:2009,Liu:2014}.

Equation~\eqref{lambda} is one of the central results of our paper. It presents the microscopic expression for the valley-mixing constant $\lambda$, which is directly proportional to the band offset at the heterojunction $V_0 = |V_+ - V_-|$. Hence, one can choose an appropriate pair of monolayers to enhance or suppress the intervalley mixing.
For a typical lattice constant $a_0=3.2$~\AA~\cite{Kormanyos:2015} and the band offset $V_0=200$~meV (corresponding to a WSe$_2$/MoS$_2$ heteropair)~\cite{Guo:2016,Davies:2021}, we obtain significant valley-mixing strength $|\lambda|\sim0.2$~eV$\cdot$\AA. 

\begin{figure}[htpb]
\begin{center}
  \includegraphics[width=0.95\linewidth]{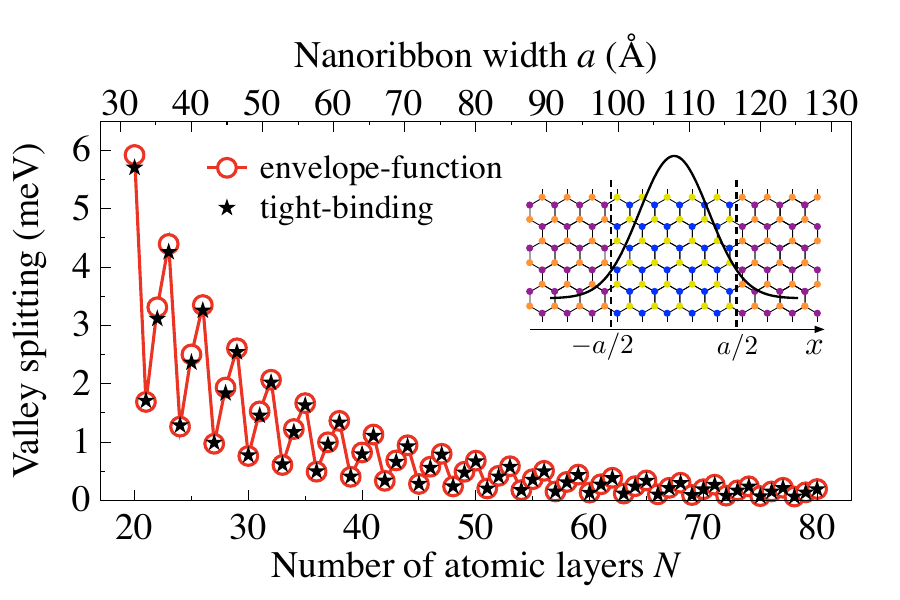}
 \end{center}
 \caption{\label{fig:valley_split} Valley splitting of the ground electron state in a nonribbon formed by a double armchair heterojunction shown in the inset. The black stars show the result of the TB calculation and red circles are calculated using the envelope-function approach, Eq.~\eqref{delta_val}.
 }
\end{figure}

A characteristic fingerprint of the intervalley mixing is the valley splitting of the localized states. It can be calculated for an electron in the symmetric nanoribbon formed by a double heterojunction shown in the inset in Fig.~\ref{fig:valley_split}. This structure is a two dimensional analog of a quantum well.
In the envelope function approximation and neglecting for simplicity the spin splitting of the conduction band, the valley splitting of the ground state is found considering Eq.~\eqref{Hval} as a perturbation and reads~\cite{Nestoklon:2006}
\begin{equation}
\label{delta_val}
\Delta_{\rm val} = 4|\lambda\cos (Ka + \varphi)| |f(a/2)|^2\:,
\end{equation}
where $a$ is the distance between the heterojunctions, $f(a/2)$ is the amplitude of the wave function in a one dimensional quantum well with the depth $V_0$ and width $a$ at the well boundary, and $\varphi$ is related to the argument of the complex parameter $\lambda$.

On the other hand, the valley splitting can be calculated directly from the TB model of a double heterojunction as an energy difference between the two lowest electron states. We set the width of the barrier layers large enough to ensure that the electron wave function does not reach the nanoribbon edges, which can also contribute to the valley splitting (see, e.g., Ref.~\cite{Avalos-Ovando:2019}).
The result is shown by black stars in Fig.~\ref{fig:valley_split} as a function of the width $a$. 
In the calculations, we use $m_c=0.5m_0$ (corresponding to $t=990$~meV) and $V_0 = 0.2t$ and obtain $\Delta_{\rm val}$ in the range of a few millielectronvolts. Similarly to previously studied nanostructures~\cite{Brey:2006, Liu:2014}, the valley splitting oscillates as a function of the number of atoms inside the quantum well $N=2a/a_0$ with the period $\Delta N = 3$ due to the trigonal symmetry of the lattice. The red circles in the same figure are plotted after analytical Eqs.~\eqref{delta_val} and~\eqref{lambda} and $\varphi=-0.4\pi$.
With increase of $V_0$, the valley splitting increases (not shown) and saturates at $V_0\gtrsim t$, because at large $V_0$ the electron wave function does not penetrate into the barrier layers. In this case, the considered structure is equivalent to a strip of the width $a$ in vacuum, and the valley splitting is induced by the edges of the strip.

The excellent agreement between Eq.~\eqref{delta_val} and microscopic calculations shown in Fig.~\ref{fig:valley_split} illustrates the strength of our analytical derivation of the valley-mixing constant from the TB model.

Note that if the interface is imperfect and $x_{\rm int}$ fluctuates along the $y$ direction, the exponent in Eq.~\eqref{Hval} changes abruptly. This can result in the suppression of the valley mixing if the electron localization length is longer than the correlation length of the interface fluctuations.


\section{Intervalley Zeeman effect} \label{sec:Zeeman}

\begin{figure}[htpb]
\begin{center}
  \includegraphics[width=0.95\linewidth]{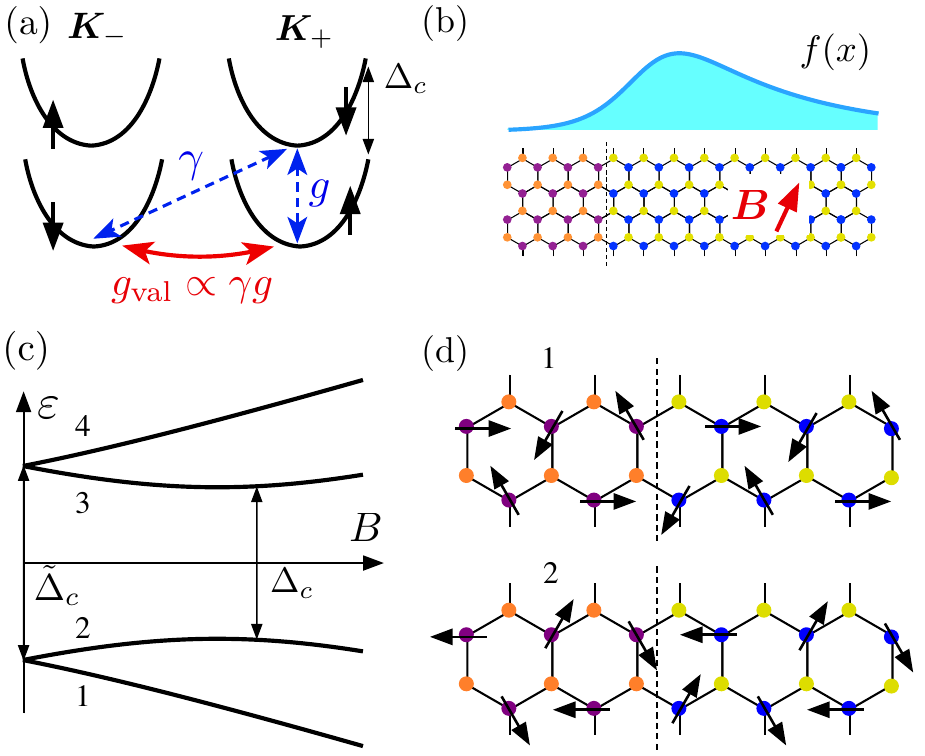}
 \end{center}
  \caption{\label{fig:Zeeman} Intervalley Zeeman effect in a lateral hetrojunction. (a) The scheme of the conduction-band states split by the spin-orbit interaction and mixed by an in-plane magnetic field ($\propto g$) and intervalley coupling ($\propto \gamma$) shown by blue arrows. The resulting intervalley Zeeman coupling (red arrows) is described by the intervalley $g$-factor $g_{\rm val} \propto \gamma g$. (b) Sketch of the electron envelope wave function at the heterojunction. (c) Fine structure of the electron states as a function of magnetic field calculated after Eq.~\eqref{eps} for $|\gamma|/\Delta_c = 0.4$. (d) Spin polarization of an electron in the states 1 and 2 in the vicinity of the heterojunction. 
}
\end{figure}

In a homogeneous TMDC monolayer, in-plane magnetic field $\bm B = (B_x, B_y)$ does not mix different valley states because of the translational invariance of the system. An atomically sharp lateral heterojunction breaks this symmetry and a finite intervalley $g$-factor appears mediated by the valley mixing described in the previous section. 
The mechanism of such an intervalley Zeeman effect is schematically shown in Fig.~\ref{fig:Zeeman}(a). 
Magnetic field couples electron states with opposite spins inside one valley, whereas interface-induced valley mixing couples the same-spin states in different valleys. As a result of this two-stage process, the Kramers-degenerate states with opposite spin and valley indices become coupled.

To describe this effect, we consider an electron localized in the vicinity of a heterojunction, as shown in Fig.~\ref{fig:Zeeman}(b). The localization may be provided by an electrostatic potential or Coulomb attraction to a hole, as in an interface exciton discussed in Sec.~\ref{sec:exc}. The electron fine structure is described by the spin-valley Hamiltonian
\begin{equation}
  \label{HB_valw}
  \mathcal H =-\frac{\Delta_c}{2}\sigma_z\tau_z + \frac12 g\mu_0 \bm\sigma \cdot \bm B+\gamma\tau_++\gamma^*\tau_-,
\end{equation}
where $\Delta_c$ is the spin splitting of the conduction band, $\bm\sigma$ is the vector composed of the spin Pauli matrices, $g$ is the intravalley $g$-factor ($g\approx2$), $\mu_0$ is the Bohr magneton, and
\begin{equation}
  \label{gamma}
  \gamma=\lambda\e^{-2\i Kx_{\rm int}}|f(x_{\rm int})|^2
\end{equation}
is the valley-mixing matrix element with $f(x)$ being the smooth envelope of the localized electron wave function.

The four eigen energies of the Hamiltonian~\eqref{HB_valw} are
\begin{eqnarray}
  \label{eps}
  \varepsilon_{1,2} &=& -\sqrt{\frac{\Delta_c^2}{4} + \left( |\gamma| \pm \frac12 g \mu_0 B \right)^2}\:, \nonumber \\
  \varepsilon_{3,4} &=& \sqrt{\frac{\Delta_c^2}{4} + \left( |\gamma| \pm \frac12 g \mu_0 B \right)^2}\:.
\end{eqnarray}
Their dependence on the magnetic field is illustrated in Fig.~\ref{fig:Zeeman}(c). From Eq.~\eqref{eps} one can see that at zero magnetic field, the intervalley mixing renormalizes the conduction band splitting to $\tilde{\Delta}_c = \sqrt{\Delta_c^2 + 4|\gamma|^2}$. 
Then, a small magnetic field leads to the splitting of the two renormalized subbands by $g_{\rm val} \mu_0 B$ with the effective intervalley $g$-factor:
\begin{equation}
\label{gval}
g_{\rm val} = \frac{2|\gamma| g}{\sqrt{\Delta_c^2 + 4|\gamma|^2}}\:.
\end{equation}
As expected from the scheme in Fig.~\ref{fig:Zeeman}(a), the intervalley $g$-factor is proportional to the product of the valley mixing parameter $\gamma$ and the intravalley $g$-factor $g$, and vanishes if the valley mixing is absent, e.g., in a lateral heterostructure with the zigzag interface. At larger magnetic fields, the energy dispersion is governed by the interplay between the $\propto B$ and $\propto B^2$ terms, the latter resulting from the intravalley Zeeman coupling. 

At small magnetic field, $g\mu_0B \ll |\gamma|$, the Zeeman term in Eq.~\eqref{HB_valw} can be treated perturbatively. As a result, the effective Hamiltonian in the basis of the two lowest (spin-up and spin-down) states has the form
\begin{equation}
  \label{HB_val}
 \mathcal H_{B}^{\rm(val)} =  -\frac{\Delta_c}{2}-\frac{g \mu_0}{\tilde\Delta_c}\left( 
\begin{array}{cc}
0 & \gamma B_- \\
\gamma^* B_+ & 0
\end{array}
\right)\:,
\end{equation}
where $B_\pm=B_x\pm\i B_y$. The eigenstates of Eq.~\eqref{HB_val} at $B \neq 0$ represent superpositions of the states in different valleys with the equal probabilities even if $\gamma$ is small. Due to the spin-valley locking, this results in a peculiar in-plane spin polarization at the metal atoms:
\begin{eqnarray}
\label{spin}
\aver{S_x (x_{n})} &=& \pm \frac12 \cos\left[ 2K(x_n - x_{\rm int}) + \phi \right]\:, \\
\aver{S_y (x_{n})} &=& \mp \frac12 \sin\left[ 2K(x_n - x_{\rm int}) + \phi \right]\:, \nonumber
\end{eqnarray}
where $\phi$ is the phase of $\lambda B_-$ and different signs correspond to the states $j=1$ and $j=2$, respectively. These expressions describe spin rotation as a function of $x$, as shown in Fig.~\ref{fig:Zeeman}(d). The spin vector rotates between the two neighbouring metal atoms by $120^\circ$ because of the large Bloch wave vector between the two valley states (see also Ref.~\cite{Avdeev:2019} for a similar effect). Note that the spins in the two states $j=1,2$ completely compensate each other at every atom.

As shown in the previous section, similarly to $\Delta_c$, $|\gamma|$ is in the range of millielectronvolts. Thus, $g_{\rm val}$ given by Eq.~\eqref{gval} can be of the order of one, but can not exceed the intravalley $g$-factor of $2$. Importantly, the Zeeman splitting in the in-plane magnetic field allows one to extract the valley relaxation time of localized electrons by measuring the Hanle curve.

\section{Interface excitons} 
\label{sec:exc}

Now we consider the effect of valley mixing on the interface excitons localized at the lateral heterojunction. Lateral heterostructures (as well as vertical ones) typically have the type-II band alignment~\cite{Guo:2016,Davies:2021,Holler:2024}. It allows for the formation of spatially indirect excitons~\cite{Lau:2018}, where electron and hole reside mainly at complementary sides of the junction and are bound to the interface due to Coulomb attraction, as illustrated in Fig.~\ref{fig:system}(a).

In the effective mass approximation, the Hamiltonian of the orbital motion of electron and hole in exciton has the form~\cite{Berkelbach:2013,Chernikov2014,Lau:2018}
\begin{equation}
  \label{eq:H_eh}
  \mathcal H_X = \frac{p_e^2}{2m_e}+\frac{p_h^2}{2m_h}+V_e(\bm r_e)+V_h(\bm r_h)+V_{\rm RK}(|\bm r_e-\bm r_h|),
\end{equation}
where $\bm r_{e,h}$ are the two dimensional electron and hole coordinates, $\bm p_{e,h}=-\i\hbar\partial/\partial\bm r_{e,h}$ are the electron and hole momenta, $m_{e,h}$ are the effective masses (assumed to be continuous across the heterojunction), $V_{e,h}(\bm r_{e,h})$ are the conduction- and valence-band potentials, and
\begin{equation}
  V_{\rm RK}(r)=-\frac{\pi e^2}{2r_0'}\left[H_0(r/r_0')-Y_0(r/r_0')\right]
\end{equation}
is the Rytova-Keldysh potential~\cite{Rytova,Keldysh1979}, which describes the electron-hole attraction in thin films. Here, $e$ is the electron charge, $H_0$ and $Y_0$ are the Struve and Neumann functions, respectively, and $r_0' = r_0/\epsilon$, where $r_0$ is the screening length of a monolayer in vacuum and $\epsilon=(\epsilon_1+\epsilon_2)/2$ is the average dielectric constant of the media above and below the monolayer.
The screening length $r_0$ weakly depends on material~\cite{Berkelbach:2013} and we further assume it to be spatially uniform. 

To simplify the calculations and focus on the new effects related to the valley mixing, we assume that the conduction and valence band offsets between the two compounds are larger than the corresponding differences of the band gaps~\cite{Davies:2021}. It allows us to write the electron and hole potentials as
\begin{equation}
\label{VeVh}
  V_{e,h}(\bm r) = V_0\Theta(\mp x),
\end{equation}
where $\Theta(x)$ is the Heaviside step function, and $V_0$ is the band offset identical for conduction and valence bands. Moreover, we assume the electron and hole masses to be equal, $m_e = m_h$, so that the problem becomes symmetric with respect to the replacements $x_e \to -x_h$ and $x_h \to - x_e$.

In contrast to previous calculations of the interface excitons~\cite{Lau:2018,Rosati:2023}, we consider an atomically sharp interface potential, see Eq.~\eqref{VeVh}, and use the variational method to find the exciton states. We consider excitons formed by electron and hole with the wave vectors near the $\bm K_\pm$-points of the Brillouin zone.
Neglecting the valley mixing, the corresponding exciton wave functions read
\begin{multline}
\label{brightXs}
\Psi_{X}^\pm (\bm r_e, \bm r_h) = \frac{\e^{\i K_y Y}}{\sqrt{L}} \Phi(x_e, x_h, y)  
 \psi_{K\pm}^c (\bm r_e) \tilde{\psi}_{K\mp}^v (\bm r_h)\:,
\end{multline}
where
$\psi_{K\pm}^c (\bm r_e)$ and $\psi_{K\pm}^v (\bm r_h)$ are the spinor Bloch functions of the valence and conduction bands in the $\bm K_\pm$-points, $Y = (y_e + y_h)/2$ is the $y$-coordinate of the exciton center of mass, $K_y$ is the corresponding wave vector, $L$ is the normalization length, $y = y_e - y_h$, and $\Phi$ is a smooth envelope function. The tilde means that the valence-band function is taken in the hole representation~\cite{Glazov:2015}.

We look for the trial function $\Phi$ in the form
\begin{equation}
\label{PhiX}
\Phi(x_e, x_h,y) = \left(\frac{2}{\pi c^2}\right)^{1/4}\e^{-y^2/c^2} f(x_e) f(-x_h)\:,
\end{equation}
where
\begin{multline}
\label{fx}
  f(x)=\frac{2}{(a+b)\sqrt{2a+b}}\left[\left(x+\frac{a}{b}x+a\right)\e^{-x/b}\Theta(x)\right.\\\left.+a\e^{x/a}\Theta(-x)\right]
\end{multline}
with $a$, $b$, and $c$ being variational parameters. Parameter $a$ describes penetration of the electron and hole wave functions under the potential barrier formed by the band offsets. Parameter $b$ is the wave function decay length away from the heterojunction ($x>0$ for the electron and $x<0$ for the hole), whereas the last parameter $c$ is the typical distance between electron and hole along the interface. 
These parameters are determined from the global minimum of the exciton energy, calculated with the Hamiltonian~\eqref{eq:H_eh}.

\begin{figure}
  \includegraphics[width=\linewidth]{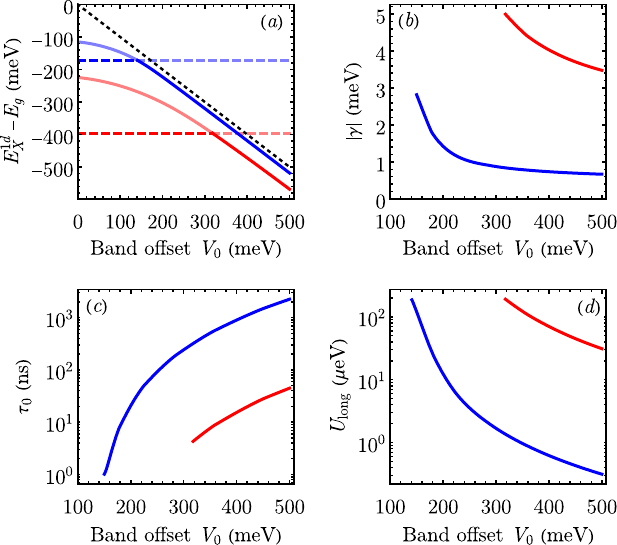}
  \caption{\label{fig:quad}
Interface excitons characteristics calculated for a lateral heterostructure on a SiO$_2$ substrate (red lines) and encapsulated in h-BN (blue lines).
    (a) Interface exciton energy counted from the band gap. Dashed lines show the energies of  free 2D excitons, black dotted line shows the $-V_0$ function. (b) Valley-mixing parameter calculated after Eq.~\eqref{eq:gamma_exc}. (c) Exciton radiative lifetime calculated after Eq.~\eqref{eq:tau}. (d)~Exciton exchange splitting calculated after Eq.~\eqref{eq:long}. 
    }
\end{figure}

The results of calculations are shown in Fig.~\ref{fig:quad}. We consider two geometries, namely,
 a lateral heterostructure supported by a SiO$_2$ substrate (with the air on the other side) and encapsulated in h-BN. We use the low-frequency dielectric constant $\epsilon_0$ for calculations of the interface excitons binding energy, since it is typically smaller than the optical phonon frequency in the surrounding media. For all other calculations, we use the high-frequency constant $\epsilon_\infty$. We take the values $\epsilon_0 = 2.45$, $\epsilon_\infty = 1.55$ for the supported and  $\epsilon_0 = 4.9$, $\epsilon_\infty = 4.45$ for the encapsulated structure~\cite{CAI2007262,PhysRevB.98.125308}. Other parameters are $m_e=m_h=0.5m_0$ and $r_0 = 4.8$~nm~\cite{Berkelbach:2013}. The binding energies of the bulk (2D) excitons are calculated with the variational function $\propto\exp\left(-\left|\bm r_e-\bm r_h\right|/a_{2d}\right)$ resulting in $E_b^{2d} \approx 396$~meV and $E_b^{2d} \approx 172$~meV for the supported and encapsulated structures, respectively~\footnote{Generally, it can be shown that the geometric average of the longitudinal and transverse dielectric constants should be used in the Rytova-Keldysh potential. For free excitons in an encapsulated monolayer, we take low frequency transverse dielectric constant of $6.71$ and high-frequency longitudinal one of $2.95$, because the obtained exciton binding energy is between the in-plane and out-of-plane longitudinal phonon energies of $200$ and $96$~meV, respectively.}.
 
The interface exciton energy $E_{X}^{1d}$ is shown by solid lines in Fig.~\ref{fig:quad}(a). To be observed in optical spectra, the interface exciton should be the lowest state, i.e. its energy should be lower than that of the bulk monolayer exciton. It happens for large enough band offsets, i.e. when $E_g-E_X^{1d} > E_b^{2d}$, where $E_g$ is the band gap, see Fig.~\ref{fig:quad}(a). The binding energy of the interface exciton $E_{b}^{1d}=E_g-E_X^{1d}-V_0$ for the supported structure is $E_{b}^{1d} \approx 80$~meV as seen from the difference of the red solid and black dotted lines, whereas for the encapsulated structure $E_{b}^{1d} \approx 20$~meV, in line with calculations of excitons localized at smooth interfaces~\cite{Rosati:2023}. For encapsulated structure and $V_0 = 220$~meV, the variational parameters in Eq.~\eqref{PhiX} are $a = 0.6$~nm, $b = 2.2$~nm and $c = 5.3$~nm corresponding to the static exciton dipole moment $d_{eh}/e \approx 5.6$~nm. Such a large value of $d_{eh}$ is comparable with the values for interface excitons bound to stacking faults in bulk semiconductors~\cite{Karin:2016,Smirnov:2018}.

Let us now examine the radiative rates of interface excitons and the fine structure of excitonic spectrum induced by the exchange interaction. It can be done by considering the self energy of 1D interface excitons, which takes into account interaction with transverse and longitudinal photons
\begin{equation}
\label{Sigma}
  \Sigma_\nu = \frac{d_{cv}^2}{\epsilon}\int \d K_x |\tilde{\Phi}(K_x)|^2\frac{K_\nu^2 - q^2}{\sqrt{K_x^2+K_y^2-q^2-\i 0}}\:.
\end{equation}
Here, $\nu = x,y$ describes the orientation of the interband optical dipole moment of the exciton $\bm d_{cv} = d_{cv} \bm e_\nu$, $\bm e_\nu$ is the unit vector parallel ($\nu = y$) or normal ($\nu = x$) to the interface, 
 $q = \sqrt{\epsilon}\omega/c$ is the photon wave vector, $\omega = E_X^{1d}/\hbar$, and
\begin{equation}
  \tilde{\Phi}(K)=\int\Phi(x,x,0)\e^{-\i Kx}\d x
\end{equation}
is the Fourier transform of the exciton wave function~\eqref{PhiX} at the same electron and hole coordinates. Equation~\eqref{Sigma} is obtained by averaging the self energy of the bulk (2D)  exciton~\cite{Gupalov:1998,Glazov:2015,Iakovlev:2024} with the 1D exciton envelope function. The exciton states active in linear polarizations with $\bm d_{cv} \parallel x$ and $\bm d_{cv} \parallel y$ are $\Psi_X^x = (\Psi_X^+ + \Psi_X^-)/\sqrt{2}$ and $\Psi_X^y = (\Psi_X^+ - \Psi_X^-)/\sqrt{2}$, respectively, with $\Psi_X^\pm$ given by Eq.~\eqref{brightXs}.

The exciton radiative lifetime $\tau_{0\nu}$ is given by the imaginary part of the self energy~\eqref{Sigma}: $\tau_{0\nu}^{-1} = -2 \Im\Sigma_\nu/\hbar$, resulting from integration over $K_x$ lying inside the light cone, $K_x^2 + K_y^2 < q^2$. Keeping in mind that the size of exciton in the $x$-direction is much smaller than the wavelength of radiation, we obtain
\begin{equation}
\label{eq:tau0_par_per}
    \frac{1}{\tau_{0x}} = \frac{1}{2 \tau_0} \left(1 + \frac{K_y^2}{q^2} \right)\:,~~ \frac{1}{\tau_{0y}} = \frac{1}{\tau_0} \left(1 - \frac{K_y^2}{q^2} \right)
  \end{equation}
for excitons polarized perpendicular and along to the interface, respectively, where the characteristic lifetime $\tau_0$ is defined as~\cite{ivchenko1992light}
\begin{equation}
  \label{eq:tau}
  \frac{1}{\tau_0}=\frac{2\pi q_0^2}{\hbar} d_{cv}^2 |\tilde{\Phi}(0)|^2\:,
\end{equation}
and $q_0=\omega/c$. Using the relation between the dipole moment $d_{cv}$ and the radiative lifetime $\tau_0^{2d}$ of the free 2D exciton~\cite{Glazov:2015} and taking into account, that typically $b \gg a$, we obtain
\begin{equation}
  \label{eq:tau2}
  \frac{\tau_0^{2d}}{\tau_0} \approx 64 \sqrt{2 \pi}  \frac{q a_{2d}^2 a^6}{c b^6} \:.
\end{equation}

The interface exciton radiative lifetime $\tau_0$ given by Eq.~\eqref{eq:tau2} is large as compared to $\tau_0^{2d}$ due to (i) small overlap between electron and hole given by the ratio $(a/b)^6 \ll 1$ and (ii) small wave vector of light $q a_{2d} \ll 1$. The dependence of $\tau_0$ on the band offset is shown in Fig.~\ref{fig:quad}(c). The curves are calculated after Eq.~\eqref{eq:tau} with $d_{cv}/e \approx 0.1$~nm, which corresponds to $\tau_0^{2d} = 1$~ps in a suspended monolayer ($\epsilon = 1$). It is seen that $\tau_0$ lies in a several~ns to 1~$\mu$s range. Interestingly, the radiative lifetime depends on exciton polarization and wave vector, see Eq.~\eqref{eq:tau0_par_per}. Such a dependence is also predicted for free 1D excitons in quantum wires~\cite{Citrin:1992}. Note that the polarization dependence vanishes for excitons localized at interface imperfections in $y$-direction if the localization length is smaller than the light wavelength and in an ensemble of excitons at the thermal equilibrium. In these cases, averages of $\tau_{0x}^{-1}$ and $\tau_{0y}^{-1}$ over $K_y$ are equal.

The long-range exchange splitting of bright exciton doublet is determined by the real part of the self energy~\eqref{Sigma}. For $K_y = 0$, we find that the exchange interaction of the exciton polarized along the interface vanishes, while the energy of another exciton is increased by
\begin{equation}
  \label{eq:long}
  U_{\rm long} = \frac{d_{cv}^2}{\epsilon}\int \d K_x |\tilde{\Phi}(K_x)|^2 |K_x|\:,
\end{equation}
where we neglected $q$. The exchange splitting is shown in Fig.~\ref{fig:quad}(d) and lies in the $\mu$eV range.

The above calculations are relevant for arbitrary crystallographic orientation of the heterointerface. At the armchair interface,
due to the valley mixing, the electron component of the exciton wave function is, in fact, a linear combination of electron states in $\bm K_+$ and $\bm K_-$ valleys, see Eq.~\eqref{envelope} and Fig.~\ref{fig:Zeeman}(a). Note that we neglect the valley mixing of holes, since it is suppressed by the large spin-orbit splitting $\Delta_v$ of the valence band~\cite{Liu:2014}. The valley-mixing constant $\gamma$ for the interface exciton is then given by Eq.~\eqref{gamma} with the electron envelope function $f(x_e)$ and reads
\begin{equation}
  \label{eq:gamma_exc}
  \left|\gamma\right|=\frac{V_0a_0}{2\sqrt{3}}f^2(0).
\end{equation}
Here, $f(x)$ is given by Eq.~\eqref{fx}, and we used the microscopic expression~\eqref{lambda} for $\lambda$. $|\gamma|$ as a function of $V_0$ is shown in Fig.~\ref{fig:quad}(b).  This dependence saturates at large $V_0$, because the decrease of the wave function at the interface is compensated by the increase of the mixing strength $|\lambda|$. Note that large, $\sim 1$~meV, values of $|\gamma|$ as compared to TMDC quantum dots are due to the small size of interface excitons.

A static electric field applied across the interface can strongly change the overlap of electron and hole wave functions. This can be used for the fine tuning of all the presented parameters of interface excitons.

\section{Polarized luminescence of interface excitons}
\label{sec:PL}

\begin{figure}
  \includegraphics[width=0.9\linewidth]{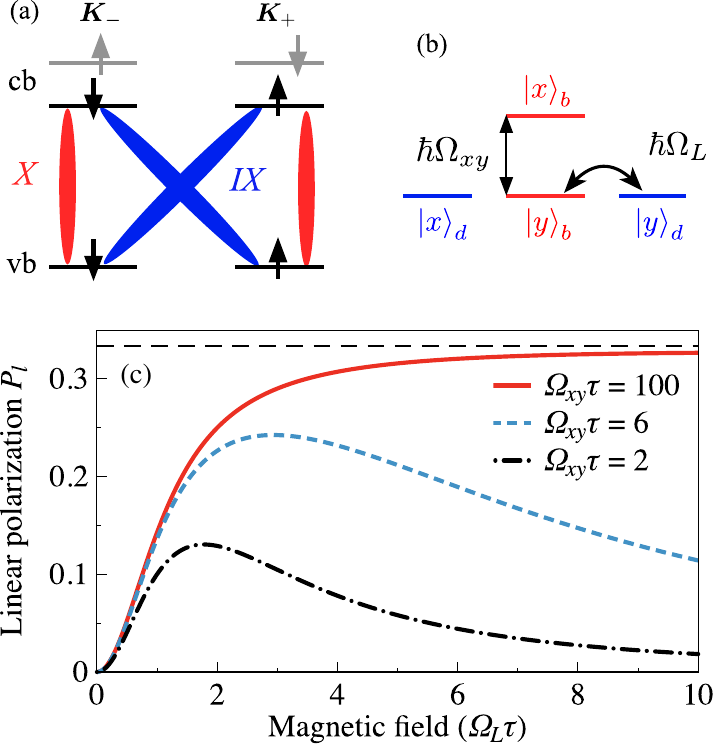}
  \caption{\label{fig:exciton_mixing} Linear polarization of interface excitons. (a) Lowest intravalley (X) and intervalley (IX) states of interface excitons. (b) Fine structure of exciton states described by
   Eq.~\eqref{eq:H_X}. (c) Degree of linear polarization of the interface excitons PL under unpolarized excitation. The curves are calculated after Eq.~\eqref{linear} for different exchange splittings given in the caption.   
   }
\end{figure}

The valley mixing of interface excitons manifests itself in the polarized photoluminescence. To demonstrate this, we consider the lowest electron subband (assuming $\Delta_c > 0$) and the topmost valence subband, see Fig.~\ref{fig:exciton_mixing}(a). They give rise to four types of excitons: two bright (direct in $\bm k$-space) and two dark (momentum indirect), which we denote as $\left| \pm \right \rangle_b$ and $\left| \pm \right \rangle_d$, respectively, with $\pm$ corresponding to the electron in the $\bm K_\pm$ valley. According to Eqs.~\eqref{Sigma} and \eqref{eq:long}, the bright excitons split into $\left| x \right \rangle_b = (\left| + \right \rangle_b + \left| - \right \rangle_b)/\sqrt{2}$ and $\left| y \right \rangle_b = (\left| + \right \rangle_b - \left| - \right \rangle_b)/\sqrt{2}$ states, which are optically active in polarizations perpendicular and parallel to interface, respectively, see Fig.~\ref{fig:exciton_mixing}(b).

Application of the in-plane magnetic field leads to the mixing of intravalley and intervalley excitons due to the intervalley electron $g$-factor introduced in Sec.~\ref{sec:Zeeman}. Making use of the Zeeman Hamiltonian~\eqref{HB_val} and a proper choice of the phases of the wave functions, the effective exciton Hamitlonian in the basis of states $\{\left| x \right \rangle_b, \left| y \right \rangle_b, \left| y \right \rangle_d , \left| x \right \rangle_d \}$ at $K_y=0$ takes the form
\begin{equation}
  \label{eq:H_X}
\mathcal H_X = \frac{\hbar}{2} \left( 
\begin{array}{cccc}
2\Omega_{xy} & 0 & 0 & \Omega_L\\
0 & 0 & \Omega_L & 0\\
0 & \Omega_L & 0 & 0\\
\Omega_L & 0 & 0 & 0
\end{array}
\right)\:,
\end{equation}
where $\Omega_{xy} = U_{\rm long}/\hbar$ is the frequency associated with the fine structure splitting, $\Omega_L = g_{\rm val} \mu_0 B/\hbar$ is the Larmor frequency, and $\left| x, y \right \rangle_d = (\left| + \right \rangle_d \pm \left| - \right \rangle_d)/\sqrt{2}$.

Coherent exciton dynamics is described by the Lindblad equation for the density matrix $\rho$:
\begin{equation}
\label{rho}
\frac{\d\rho}{\d t} = \frac{\i}{\hbar} \left[ \rho, \mathcal H_X \right] + G - R\{\rho\}\:,
\end{equation}
where $G$ and $R$ describe exciton generation and recombination, respectively. The former, under resonant optical excitation, has four nonzero components:
\begin{equation}
G_{11} \propto \frac{1 + P_l^{(0)}}{2}\:,~G_{22} \propto \frac{1 - P_l^{(0)}}{2}\:,~G_{12}  \propto \frac{P_c^{(0)} + \i P_{l'}^{(0)}}{2}\:,
\end{equation}
with a common factor, and $G_{21} = G_{12}^*$. Here, $P_l^{(0)}$, $P_{l'}^{(0)}$, and $P_c^{(0)}$ are the Stokes parameters of the exciting light, horizontal (vertical), diagonal and circular in the $x,y$ axes, respectively.
Because of the slow radiative recombination of the interface excitons, see Fig.~\ref{fig:quad}(c), we assume that the recombination is mostly nonradiative $R\{\rho\}=-\rho/\tau$ with the nonradiative decay time $\tau \ll \tau_0$.

In the steady state, the time derivative in Eq.~\eqref{rho} vanishes, and the solution yields the degree of linear polarization of PL in the $x,y$ axes $P_l = (\rho_{11} - \rho_{22})/(\rho_{11} + \rho_{22})$. Quite naturally, for linearly polarized excitation along or perpendicular to the interface, ($P_l^{(0)} = \mp 1$, respectively), PL has the same polarization, $P_l = P_l^{(0)}$. More curiously, for other polarizations of the excitation, $P_{l'}^{(0)} = \pm 1$ or $P_c^{(0)} = \pm 1$, and even for unpolarized excitation, $P_l^{(0)} = P_{l'}^{(0)} = P_c^{(0)} = 0$, we find that PL is still linearly polarized across the interface:
\begin{equation}
\label{linear}
P_l = \frac{\Omega_L^2 \Omega_{xy}^2 \tau^4}{4 + 2 \Omega_L^4 \tau^4 + 3 \Omega_L^2 \Omega_{xy}^2 \tau^4 + (6 \Omega_L^2 + 4 \Omega_{xy}^2) \tau^2}\:.
\end{equation}
This is a consequence of the mixing of intra- and intervalley excitons by magnetic field.

The magnetic field dependence of $P_l$ is plotted in Fig.~\ref{fig:exciton_mixing}(c) for different values of the exchange splitting $\Omega_{xy}$. Generally, it is a nonmonotonous function, which approaches $1/3$ for $\Omega_{xy}\tau\gg1$.

The origin of the linearly polarized PL of interface excitons under unpolarized optical excitation is illustrated in Fig.~\ref{fig:exciton_mixing}(b). The in-plane magnetic field mixes the bright $\left| y \right \rangle_b$ state with the dark $\left| y \right \rangle_d$ state, hence, suppressing $\left| y \right \rangle_b$  luminescence up to two times. As a result, PL is polarized along the $x$ axis by up to $33\%$. However, if magnetic field becomes sufficiently strong, $\Omega_L\ge\Omega_{xy}$, the state $\left| x \right \rangle_b$ gets coupled with $\left| x \right \rangle_d$ resulting in the decrease of the polarization degree.  For a typical exchange splitting $U_{\rm long} = 10~\mu$eV, the largest linear polarization of $1/3$ can be reached at $\tau \gg 0.1$~ns.

\section{Conclusion}

To summarize, we have developed a microscopic theory of the valley mixing at lateral heterojunctions and have described its consequences on the dynamics of localized electrons and interface excitons.
Using the TB model, we have obtained an analytical expression for the valley mixing constant resulting in a value of $\sim 0.2$~eV$\cdot$\AA~for typical heteropairs. For interface excitons, this yields the intervalley coupling matrix element of $\sim 1$~meV comparable to the spin splitting of the conduction band. We have shown that the valley mixing results in the coupling of Kramers degenerate electron states by the in-plane magnetic field, i.e.  the intervalley Zeeman effect, forbidden in homogeneous TMDC monolayers. We have found that the interplay of the valley mixing and long-range exchange interaction leads to the linear polarization of exciton photoluminescence along the armchair heterojunction with the degree of polarization up to $1/3$ under unpolarized excitation. 

\acknowledgements

We thank \href{https://www.graphene.manchester.ac.uk/research/people/vladimir-falko/}{V. I. Fal'ko} and \href{http://www.ioffe.ru/coherent/index.html/Coherent/Staff.html}{M. M. Glazov} for fruitful discussions. D.S.S. acknowledges financial support from the Russian Science Foundation Grant No. 23-12-00142 and the Foundation for the Advancement of Theoretical Physics and Mathematics ``BASIS.''

\renewcommand{\i}{\ifr}

%
\end{document}